# Direct acceleration of an electron in infinite vacuum by a pulsed radially-polarized laser beam

Liang Jie Wong* and Franz X. Kärtner

*Department of Electrical Engineering and Computer Science and Research Laboratory of Electronics, Massachusetts Institute of Technology, 77 Massachusetts Avenue, Cambridge, MA, 02139, USA*
*ljwong@mit.edu*

**Abstract:** We study the direct acceleration of a free electron in infinite vacuum along the axis of a pulsed radially-polarized laser beam. We find that net energy transfer from laser pulse to electron is maximized with the tightest focusing. We show that the net energy gain of an electron initially moving at a relativistic velocity may exceed more than half the theoretical limit of energy transfer, which is not possible with an initially stationary electron in the parameter space studied. We determine and analyze the power scaling of maximum energy gain, extending our study to include a relatively unexplored regime of low powers and revealing that substantial acceleration is already possible without the use of petawatt peak-power laser technology.

## 1. Introduction

Electron acceleration is a rapidly-advancing field of scientific research with widespread applications in industry and medicine [1-3]. With the invention of chirped pulse amplification [4], by which lasers with petawatt peak powers [5] and ultrahigh intensities [6] have been realized, there has been growing interest in laser-driven electron acceleration schemes due to their potential to offer compact and low-cost setups through high accelerator gradients [1, 7, 8]. The use of a plasma medium [7] is an attractive way of achieving laser-driven electron acceleration, but faces problems like the inherent instability of laser-plasma interactions. This has prompted the investigation of laser-driven electron acceleration in vacuum, which takes place primarily through either the ponderomotive force associated with the transverse electric and magnetic field components (ponderomotive acceleration) [9-20], or the force exerted by the longitudinal electric field component (direct acceleration) [21-27].

Among the variety of ponderomotive acceleration schemes conceived are inverse free electron laser (IFEL) acceleration [9, 10]; vacuum beat wave acceleration [12], in which the wiggler field of the IFEL is simply replaced by a second laser; high-intensity ponderomotive scattering [13-15], in which the electron is scattered away from the laser focus with a high escape energy; the capture and acceleration scenario [17], in which relativistic electrons are injected at an angle into the laser focus; and ionization of highly-charged ions near the laser pulse peak [19]. Experiments [10, 15, 20] have demonstrated that ponderomotive acceleration may be achieved in reality. A number of direct acceleration schemes [21, 22] that involve terminating the laser field before the accelerated electron starts losing energy to the field have also been proposed. However, the presence of optical components near the laser focus limits the laser field intensity that may be used due to material damage concerns. The use of a gas medium [28] to replace the additional optical components in schemes like [22] only introduces new limitations associated with ionization of the gas.

Direct acceleration of electrons in infinite vacuum by a pulsed radially-polarized laser beam is particularly attractive because such a scheme places no limit on the laser field intensity that may be used. The linear nature of direct acceleration also leads to low radiative losses, as calculations in [26] confirm. In addition, studies [23, 24] have shown that the off-axis radial electric and azimuthal magnetic field components of the radially-polarized laser help confine electrons to the vicinity of the beam axis, favoring the production of mono-energetic and well-collimated electron beams. Three dimensional Particle-in-Cell simulations of electron acceleration in vacuum [27] comparing the performance of a pulsed Gaussian laser beam to that of a pulsed radially-polarized laser beam have demonstrated the superiority of the latter in terms of electron beam quality and maximum energy gain.

In this paper, we study the direct acceleration of a free electron in infinite vacuum along the axis of a pulsed radially-polarized laser beam. All optimizations are carried out in view of maximizing net energy transfer from laser pulse to electron. We begin by studying the initially stationary electron. There has been some interest [23-27] in the scenario of electrons born (for instance, by ionization) in the path of the laser pulse, and a recent study by Fortin et. al. [25] showed that an electron can reach the high-intensity cycles of the pulse without having been released by photoionization near the pulse peak. The study also concluded that the optimal beam waist at petawatt peak powers lies well within the paraxial wave regime. The latter conclusion, however, is true only for an initially stationary electron required to start at the laser focus. We show that after including the electron's initial position in the optimization space, we in fact achieve maximum acceleration with the most tightly-focused laser.

To the best of our knowledge, there has been no extensive study on our subject for electrons with non-zero initial velocities. These electrons (which we call "pre-accelerated electrons") are injected into the laser beam ahead of the pulse and may be the output of a preceding acceleration stage. We show that net energy gain can be much greater for a pre-accelerated electron than for an initially stationary one. In particular, the net energy gain of an initially relativistic electron may exceed more than half the theoretical energy gain limit (derived in [25]), which is not possible with an initially stationary electron in the parameter space studied. The *de facto* energy gain limit (of half the theoretical energy gain limit) argued by Fortin et. al. [25] for the initially stationary electron may thus be surpassed with the pre-accelerated electron. Based on our simulation results, we also hypothesize that substantial electron acceleration cannot be achieved if the electron's initial kinetic energy greatly exceeds the laser's theoretical gain limit.

To the best of our knowledge, there has also been no extensive study on our subject for laser peak powers below 100 TW; most studies focus on petawatt powers and beyond. By extending our parameter space to include powers as low as 5 TW, we show that substantial acceleration can already be achieved with laser peak powers of a few terawatts. In particular, we give an example in which a 5 TW pulse, either 7.5 fs or 15 fs in pulse duration, accelerates an electron from a kinetic energy of 10 MeV to a kinetic energy of about 50 MeV; and another example in which a two-stage accelerator employing a 10 TW, 10 fs pulse in each stage accelerates an initially stationary electron to a final kinetic energy of about 36 MeV. These electron energies are already sufficient for applications like the production of hard X-rays via inverse Compton scattering [29].

The paper is organized as follows: In Section 2, we discuss the theoretical and technical aspects of our simulations; in Section 3, we study the acceleration of an initially stationary electron; in Section 4, we study the acceleration of a pre-accelerated electron; in Section 5, we conclude with a summary of our findings.

## 2. Theory of direct acceleration by a pulsed radially-polarized laser beam

*2.1 Overview*

The physical scenario we study is the following: A free electron, initially at rest or moving in field-free vacuum, is overtaken by the pulse of a radially-polarized laser beam that exchanges energy with the electron purely via the laser's on-axis, longitudinal electric field (i.e. via direct acceleration). The pulse eventually overtakes the electron, leaving the electron once again in field-free vacuum, with a velocity generally different from what it had before. The free electron may have been introduced either by ionization of a target in the path of the pulse, as in [27], or by a preceding acceleration stage. To compute the net energy gain of the electron, we need a description of the laser pulse and equations to model the electron's motion.

*2.2 Description of a pulsed radially-polarized laser beam*

Using the method of [30], we may derive the electric field $\vec{E}$ and magnetic flux density $\vec{B}$ for a pulsed radially-polarized laser beam in vacuum under the paraxial wave approximation:

$$\vec{E}(r,z,t) = \mathrm{Re}\left\{\tilde{E}(r,z)e^{j(\xi+\psi_0)}\mathrm{sech}\left(\frac{\xi+kz_i}{\xi_0}\right)\right\}, \quad \vec{B}(r,z,t) = \hat{\phi}\frac{1}{c}\vec{E}(r,z,t)\cdot\hat{r}, \qquad (1)$$

where

$$\tilde{E}(r,z) \equiv f^2\rho e^{-f\rho^2}\sqrt{\frac{8\eta_0 P}{\pi w_0^2}}\left[\hat{r}-\hat{z}\frac{2j}{kr}\left(1-f\rho^2\right)\right], \quad \tilde{B}(r,z) \equiv \hat{\phi}\frac{1}{c}\tilde{E}(r,z)\cdot\hat{r}, \qquad (2)$$

$r,\phi,z$ are the cylindrical coordinates and $\hat{r},\hat{\phi},\hat{z}$ the corresponding unit vectors; $j \equiv \sqrt{-1}$; $f \equiv j/(j+(z/z_0))$; $\rho \equiv r/w_0$; $\xi \equiv \omega t - kz$; $z_0 \equiv \pi w_0^2/\lambda$ is the Rayleigh range; $w_0$ is the beam waist radius; $\lambda$ is the carrier wavelength (i.e. the central wavelength of the pulse); $k \equiv 2\pi/\lambda$; $\omega = kc$ is the angular carrier frequency; $\eta_0 \cong 120\pi\,\Omega$ is the vacuum wave impedance; $c$ is the speed of light in vacuum; $z_i$ is the pulse's initial position; $\psi_0$ is the carrier phase constant; $\xi_0$ is a parameter related to the pulse duration; $P$ is the peak power of the pulse:

$$P = \frac{1}{2\mu_0}\int_0^\infty dr\, 2\pi r\, \mathrm{Re}\{\tilde{E}(r,0)\times\tilde{B}^*(r,0)\cdot\hat{z}\} \qquad (3)$$

where $\mu_0$ is the permeability of free space. By choosing values of $\xi_0$ such that the time variation of the sech pulse envelope is large compared to the time variation of the carrier and using Eq. (3), we may compute the pulse energy $E_{pulse}$ as

$$E_{pulse} = \frac{1}{\mu_0}\int_{-\infty}^\infty dt\int_0^\infty dr\, 2\pi r\vec{E}(r,0,t)\times\vec{B}(r,0,t)\cdot\hat{z} \approx P\int_{-\infty}^\infty dt\,\mathrm{sech}^2\left(\frac{\omega t+kz_i}{\xi_0}\right) = P\frac{2\xi_0}{\omega} \qquad (4)$$

We have chosen to model our pulse with a sech envelope because this allows Eq. (1) to satisfy the Maxwell equations in the paraxial wave approximation for $\xi_0 \gg 1$. As shown in [30], the same cannot be said for other choices of pulse shapes. In particular, using a Gaussian pulse $\exp(-(\xi+kz_i)^2/\xi_0^2)$, instead of $\text{sech}((\xi+kz_i)/\xi_0)$, would cause Eq. (1) to violate the Maxwell equations at large values of $(\xi+kz_i)$ (i.e. at the tails of the pulse). However, as will be seen in the next section, we are able to reproduce the results of [25] – which used a Gaussian pulse – with our model, showing that the former approach does not suffer much in accuracy in the parameter space of [25]. This is because the electrodynamics for most cases in [25] is primarily influenced by fields close to the pulse peak, where both Gaussian and sech representations are accurate.

Following the convention of [25], we define the pulse duration $\tau$ to be the single-sided $\exp(-1)$ duration of the pulse:

$$\tau = \frac{\xi_0}{\omega}\text{sech}^{-1}(\exp(-1)). \tag{5}$$

Eq. (1) thus uniquely defines a pulsed radially-polarized laser beam after we specify six parameters: carrier wavelength $\lambda$, carrier phase constant $\psi_0$, beam waist radius $w_0$, initial pulse position $z_i$, peak power $P$ and pulse duration $\tau$. The pulse energy $E_{pulse}$ and parameter $\xi_0$ are then fixed by equations (4) and (5) respectively.

*2.3 Relativistic electrodynamics of an on-axis electron*

The electrodynamics of an electron in an electromagnetic field, ignoring radiative reaction, is described by the Newton-Lorentz equation of motion [31]

$$\frac{d\vec{p}}{dt} = \frac{d(\gamma m\vec{v})}{dt} = -e(\vec{E}+\vec{v}\times\vec{B}), \tag{6}$$

where $r,\phi,z$ in the variables of Eq. (6) now denote the coordinates of the electron's position, $m$ is the rest mass of the electron, $e$ the absolute value of its charge, $\vec{p}$ its momentum, $\vec{v}$ its velocity and $\gamma \equiv 1/\sqrt{1-\beta^2}$ is the Lorentz factor, with $\beta \equiv |\vec{\beta}|$ and $\vec{\beta}\equiv \vec{v}/c$. The total energy and kinetic energy of the electron are given by $E_t = \gamma mc^2$ and $E_K = (\gamma-1)mc^2$ respectively.

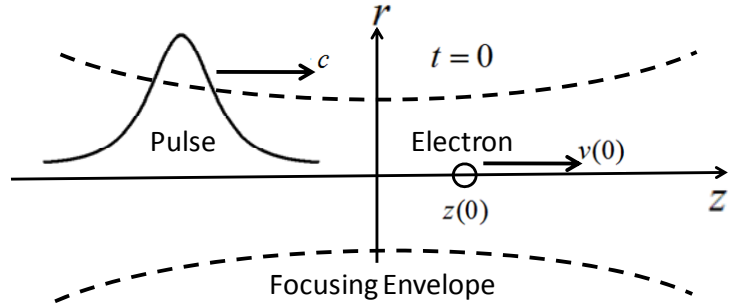

Fig. 1. Schematic of simulations at initial time.

We consider an electron initially ($t=0$) on the beam axis ($r=0$) of the laser at $z=z(0)$ (Fig. 1), moving in the longitudinal direction with velocity $\vec{v}(0)=v(0)\hat{z}$. The electron may be initially at rest ($v(0)=0$) or moving ($v(0)>0$; we do not consider $v(0)<0$). The former case is the subject of Section 3, whereas the latter is the subject of Section 4. In all cases, we are interested in the net energy the electron extracts from the laser field as the pulse propagates from a position (effectively) infinitely far behind the electron to a position (effectively) infinitely far in front of the electron. We do not limit the interaction distance by use of any additional optics. We also confine our attention to forward scattering cases (i.e. the electron's final velocity is in the direction of pulse propagation $+z$).

Setting $r = 0$ in Eq. (1), we have $\vec{E} = E_z \hat{z}$ and $\vec{B} = 0$, where

$$E_z = \left[ \frac{1/z_0}{1+(z/z_0)^2} \sqrt{\frac{8\eta_0 P}{\pi}} \right] \sin\left( \omega t - kz + 2\tan^{-1}\left(\frac{z}{z_0}\right) + \psi_0 \right) \text{sech}\left( \frac{\omega t - k(z-z_i)}{\xi_0} \right) \quad (7)$$

Eq. (7) may be seen as the product of three parts: the *field amplitude*, given by the square bracketed factor, which is a Lorentzian in $z$; the *continuous wave* (*CW*) *carrier*, given by the $\sin(\cdot)$ factor; and the *pulse envelope*, given by the $\text{sech}(\cdot)$ factor. The sign of $E_z$ is determined exclusively by that of the CW carrier. If the CW carrier is positive, meaning its argument is between 0 and $\pi$ radians, an electron traveling in the $+z$ direction is in a *decelerating cycle* and loses energy to the field. If the CW carrier is negative, meaning its argument is between $\pi$ and $2\pi$ radians, an electron traveling in the $+z$ direction is in an *accelerating cycle* and gains energy from the field. An on-axis electron with no initial transverse velocity component is confined to move along the beam axis (so $\vec{v}(t) = v(t)\hat{z} \ \forall t$, $r(t) = 0 \ \forall t$). Simplifying Eq. (6), we obtain the equations

$$\frac{d\beta}{dt} = -\frac{eE_z}{\gamma^3 mc}, \quad \frac{dz}{dt} = v = c\beta. \quad (8)$$

Eq. (8) may be solved numerically for the electron's speed, and hence its energy, at any time. To do so, however, we must first specify the laser field (by specifying $\lambda$, $\psi_0$, $w_0$, $z_i$, $P$ and $\tau$) as well as the electron's initial position $z(0)$ and speed $v(0)$. As mentioned, we always set $z_i$ such that the pulse effectively begins infinitely far behind the electron. In addition, we fix $\lambda = 0.8\,\mu m$ throughout the paper, leaving us with a total of six dimensions over which to study or optimize the problem. Although we fix $\lambda$, our results may be readily scaled to obtain the results for any $\lambda$ by nature of Eqs. (7) and (8), as we see in the next sub-section.

*2.4 Scalability of solutions to any central wavelength*

If we let $T \equiv \omega t$, $\varsigma \equiv z/z_0$ (with $\varsigma_i \equiv z_i/z_0$) and $\kappa \equiv kz_0 = 2(\pi w_0/\lambda)^2$, and apply Eq. (7), Eq. (8) may be cast in the form

$$\frac{d\beta}{dT} = -\frac{e}{\gamma^3 mc^2} \left[ \frac{1/\kappa}{1+\varsigma^2} \sqrt{\frac{8\eta_0 P}{\pi}} \right] \sin\left(T - \kappa\varsigma + 2\tan^{-1}(\varsigma) + \psi_0\right) \text{sech}\left(\frac{T - \kappa(\varsigma-\varsigma_i)}{\xi_0}\right), \quad \frac{d\varsigma}{dT} = \frac{\beta}{\kappa} \quad (9)$$

For given values of $\kappa$ and $\xi_0$, Eq. (9) is completely independent of central wavelength $\lambda$. The results for any $\lambda$ may thus be obtained from the results for $\lambda = 0.8\,\mu m$ by an appropriate scaling of beam waist $w_0$ and pulse duration $\tau$. Note that $\kappa$ determines the ratio $w_0/\lambda$ and $\xi_0$ determines the number of cycles in the pulse envelope, regardless of $\lambda$. The scaling in $t$ and $z$ does not affect the maximum energy gain, only the optimal $z_i$.

By substituting Eqs. (1) and (2) into Eq. (6) and applying $\rho \equiv r/w_0$ along with the previous normalizations, it is straightforward to generalize our conclusion and see that for given values of $\kappa$ and $\xi_0$, the electrodynamic equations are independent of $\lambda$ even for the most general case where the electron is not necessarily on the beam axis. The acceleration of an (on-axis or otherwise) electron in infinite vacuum by a pulsed radially-polarized laser beam thus depends on $\lambda$ only through $\kappa$ and $\xi_0$. An important consequence of this is that for a given peak power $P$, a larger pulse energy is required for exactly the same maximum acceleration at a larger $\lambda$ if focusing ($w_0/\lambda$) remains constant, because the number of carrier cycles in the pulse envelope must also remain constant, leading to a longer pulse.

*2.5 The theoretical energy gain limit*

The Gouy phase shift term $2\tan^{-1}(z/z_0)$ in the argument of the CW carrier in Eq. (7) prevents any particle from remaining in a single cycle indefinitely. As a result, the energy that an electron can gain from a pulsed radially-polarized laser beam has a theoretical limit $\Delta E_{\lim}$ that may be computed by considering an electron that (unrealistically) remains at the pulse peak and in one accelerating cycle from the focus to infinity (or from $-z_0$ to $z_0$, which gives the same result, just with a different $\psi_0$), as was done in [25]:

$$\Delta E_{\lim} = \int_0^\infty dz \left[ \frac{e/z_0}{1+(z/z_0)^2} \sqrt{\frac{8\eta_0 P}{\pi}} \right] \sin(2\tan^{-1}(z/z_0)) = e\sqrt{\frac{8\eta_0 P}{\pi}} \cong \sqrt{\frac{P}{[PW]}}[GeV] \quad (10)$$

Where $P/[PW]$ refers to the laser peak power in petawatts. We will find it convenient to normalize our energy gain results by $\Delta E_{\lim}$ afterwards.

*2.6 Technical aspects and validity of the simulations*

We solve Eq. (8) numerically via the Adams-Bashforth-Moulton method (*ode113* of *Matlab*). In every case, we ensure that the pulse begins so far behind the electron that the latter is initially not affected by the laser field. By this we mean that any fluctuation in the electron's energy is at first (for at least a few tens of picoseconds) below an arbitrarily small value. We also terminate our simulations only after the electron's energy has reached a steady state (equivalently, after electron position $z$ has become so large that the Lorentzian field amplitude of Eq. (7) is negligibly small).

As discussed, Eq. (1) satisfies the Maxwell equations only for sufficiently large beam waists and pulse widths. To ensure the validity of our simulations, the smallest waist and pulse duration we consider are respectively $w_0 = 2\mu m$ and $\tau = 7.5\,fs$, after the fashion of Fortin et. al. [25] and based on findings by Varin et. al. [32] that corrections to the paraxial radially-polarized laser beam are small or negligible for beam waists no smaller than $w_0 = 2\mu m$. For $\tau \geq 7.5\,fs$, $\xi_0 > 10$, which at least approximately satisfies the requirement that $\xi_0 \gg 1$.

## 3. Direct acceleration of an initially stationary electron

*3.1 Simulation Results and Analysis*

In [25], Fortin et. al. studied the case of a pulsed radially-polarized laser beam incident on an electron that was initially stationary at the laser focus. The authors concluded that, for the range of laser peak powers and pulse durations studied, the optimal laser focusing is in general not the tightest. This conclusion, however, is true only for electrons required to start at the laser focus (i.e. $z(0) = 0$). Given $P$, $\tau$ and $w_0$ in general, $z(0) = 0$ (or even slightly less than 0, as the authors suggest) is not the optimal initial position. We find after optimizing over $\psi_0$ - $w_0$ - $z(0)$ space that the optimal focusing is in fact the tightest.

In Fig. 2, we plot the maximum energy gain and optimal beam waist computed by optimizing over $\psi_0$ - $w_0$ space for $z(0) = 0$ (as in [25]). In Fig. 2(a), we also plot the maximum energy gain computed by optimizing over $\psi_0$ - $w_0$ - $z(0)$ space for $w_0 \geq 2$ μm (giving optimal $w_0 = 2$ μm). Our results for $z(0) = 0$ are clearly in good agreement with those in [25] (slight differences may be attributed to our use of a different pulse shape). We see that a substantial increase in maximum energy gain occurs after including the $z(0)$ dimension in the optimization space. In fact, 15 fs and 20 fs pulses can approximately give us the energy gain that for $z(0) = 0$ is achievable only with 7.5 fs and 10 fs pulses respectively.

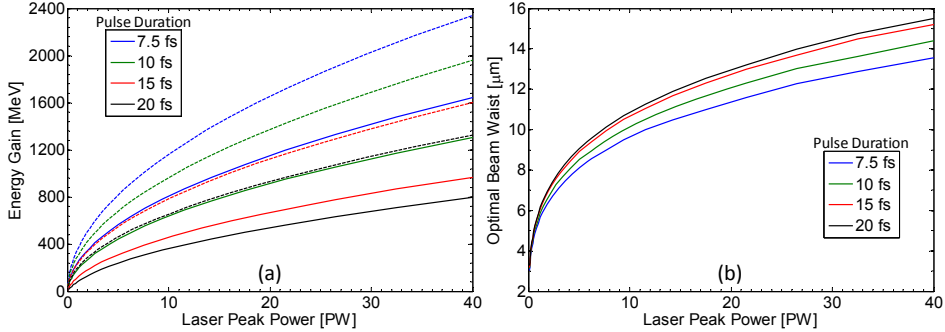

Fig. 2. (Color online) (a) Maximum energy gain and (b) corresponding optimal beam waist vs. power $P$ from 0.1 to 40 PW for various $\tau$. All solid lines correspond to $z(0) = 0$. Dashed lines correspond to optimal $z(0)$ for $w_0 = 2$ μm (optimal waist). All cases shown correspond to forward scattering of the electron.

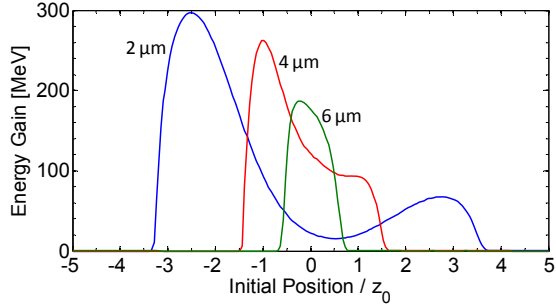

Fig. 3. (Color online) Maximum energy gain vs. normalized $z(0)$ for $P = 1$ PW, $\tau = 10$ fs and various $w_0$. All cases shown correspond to forward scattering of the electron.

To illustrate how $z(0) = 0$ is not optimal in general, the energy gain (maximized over $\psi_0$ space) as a function of $z(0)$ normalized by $z_0$ for a 1 PW, 10 fs pulse is plotted for various waists in Fig. 3. As can be seen, the optimal $z(0)$ approaches the focus as $w_0$ increases for given $P$ and $\tau$, but in general may be quite a distance behind the focus.

We would like to evaluate the power scaling characteristics for various $\tau$ and $w_0$, extending our region of study to include laser peak powers as low as 5 TW. The results of optimization over $\psi_0$ - $z(0)$ space are shown in Fig. 4. To improve readability, we have normalized the electron's maximum energy gain at each $P$ by the gain limit $\Delta E_{\text{lim}}$ (Eq. (10)), and the electron's optimal initial position by the Rayleigh range $z_0$. Note that the $w_0 = 2$ μm plots in Fig. 4(a) are just normalized versions of the dashed lines in Fig. 2(a). From Fig. 4, we observe the following trends:

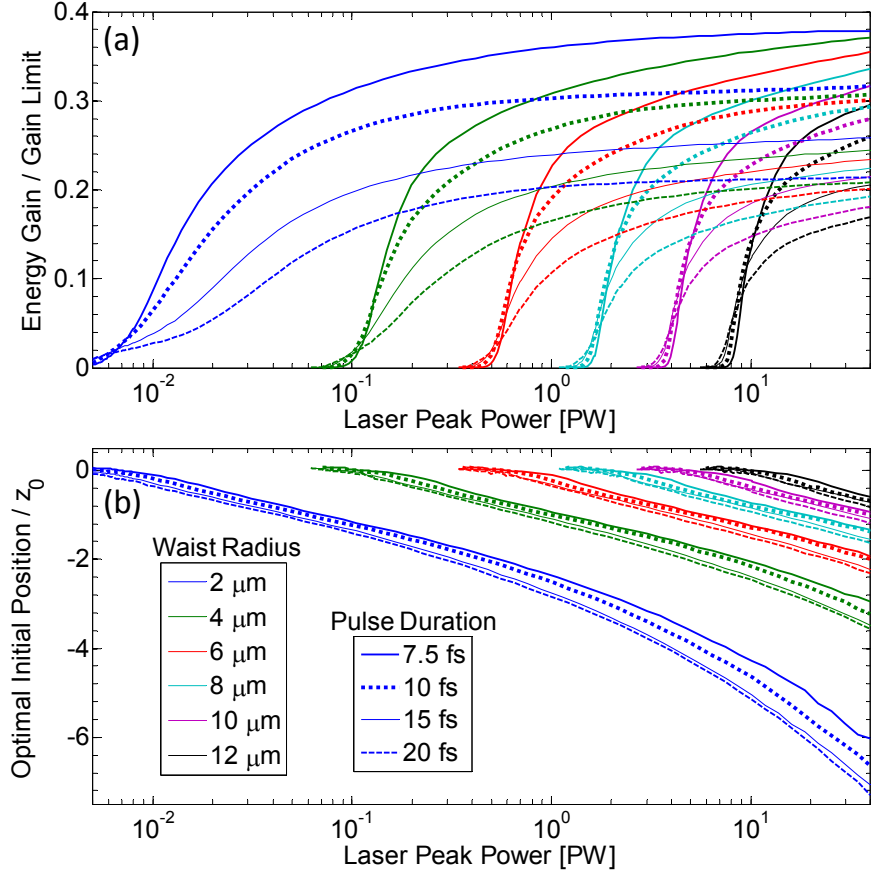

Fig. 4. (Color online) (a) Normalized maximum energy gain and (b) corresponding normalized optimal initial position vs. $P$ from 5 TW to 40 PW for various $w_0$ and $\tau$. All cases shown correspond to forward scattering of the electron. Cases of very non-relativistic final kinetic energy are not plotted to reduce clutter.

a) Given $\tau$ and $w_0$, a threshold power $P_{th}$ exists such that negligible energy gain is obtained for $P < P_{th}$. $P_{th}$ is approximately independent of $\tau$ and is approximated by the condition used in [25] to find the threshold $w_0$ for given $P$ with $z(0) = 0$:

$$a_0 \equiv \frac{e}{mc\omega z_0}\sqrt{\frac{8\eta_0 P_{th}}{\pi}} \approx 1 \qquad (11)$$

where $a_0$ is simply the normalized field amplitude of $E_z$ at the focus. As discussed in [25], Eq. (11) is motivated by the observation made in ponderomotive acceleration studies (e.g. [14]) that $a_0 \geq 1$ is required to access the relativistic regime of laser-electron interaction (except that for ponderomotive acceleration, $a_0$ is computed with the transverse rather than longitudinal field amplitude). For $w_0 = 2, 4, 6, 8, 10, 12$ μm, Eq. (11) gives $P_{th} \approx 4.163\times10^{-3}$, $6.661\times10^{-2}$, $3.372\times10^{-1}$, $1.066$, $2.602$, $5.396$ PW (4 sig. fig.) respectively, which by Fig. 4(a) are estimates accurate to well within an order of magnitude.

b) Given $\tau$ and $w_0$, energy gain (whether in MeV or normalized by $\Delta E_{\lim}$) increases with increasing $P$. That the normalized gain asymptotically approaches a constant value tells us that at $P \gg P_{th}$, the energy gain in MeV is approximately proportional to $\sqrt{P}$, a behavior that has been noted for the $z(0) = 0$ case studied in [25].

c) Given $w_0$ and $P$, energy gain increases with increasing $\tau$ up to an optimal $\tau$ and decreases as $\tau$ increases further. As the given $P$ decreases toward $P_{th}$, this optimal $\tau$ increases, showing that longer pulses are favored at lower powers. A close-up of Fig. 4(a) with energy gain in MeV is shown in Fig. 5 to illustrate this. The conclusion of [25] that a shorter pulse leads to greater net acceleration is thus not generally true.

d) Given $\tau$ and $P$, energy gain decreases with increasing $w_0$. As far as we can determine in the paraxial wave approximation, the optimal focusing for direct electron acceleration is the tightest.

e) Given $\tau$ and $w_0$, the optimal initial position becomes more negative with increasing $P$ for the vast majority of cases, especially where $P \gg P_{th}$, in Fig. 4(b). At $P \approx P_{th}$, the optimal initial position is close to the focus and may even be slightly positive. For $P \gg P_{th}$, the optimal initial position is negative and approximately proportional to $\sqrt[4]{P}$, as we have ascertained by curve-fitting.

f) Given $w_0$ and $P$, the optimal initial position becomes more negative with increasing $\tau$ for the vast majority of cases, especially where $P \gg P_{th}$, in Fig. 4(b).

g) Given $\tau$ and $P$, the optimal initial position normalized by $z_0$ becomes more negative with decreasing $w_0$.

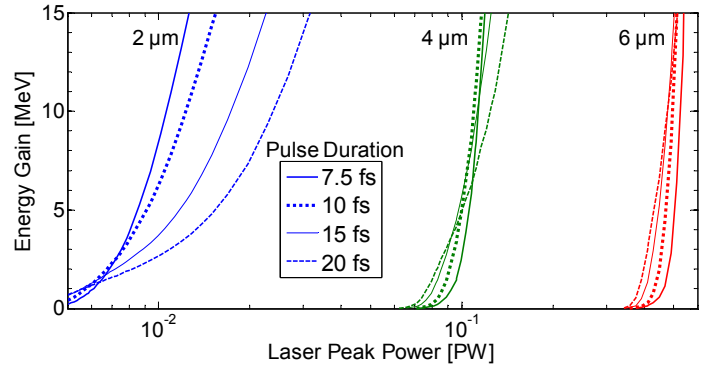

Fig. 5. (Color online) Close-up of plot of maximum energy gain vs. $P$ for various $w_0$ and $\tau$.

One may intuitively expect $z(0) = 0$ to be the optimal initial position in general since, after all, the theoretical gain limit $\Delta E_{\lim}$ was computed in Eq. (10) by assuming an electron that enters an accelerating cycle at the laser focus and staying in that cycle forever. However, an electron that starts at rest is bound to slip through a succession of accelerating and decelerating cycles before entering what is effectively its final accelerating cycle (that is, the final accelerating cycle that has any significant impact on its energy) with a velocity that is in general quite different from its initial velocity, so the relationship between $z(0)$ and the electron's final energy gain is complicated. We also note that although including the $z(0)$ dimension in the optimization space significantly increases the electron's energy gain over the $z(0) = 0$ case, the electron still extracts at best less than $\Delta E_{\lim}/2$ of energy from the pulse. In [25], it is argued that sub-cycle direct acceleration can only take place from $z > z_0$ to $\infty$, so the energy gain will always be less than $\Delta E_{\lim}/2$ for initially stationary electrons. We show in the next section that by using a pre-accelerated electron, we can make the electron enter its final accelerating cycle at a position $z < z_0$ and extract more than $\Delta E_{\lim}/2$ of energy from the pulse

## 4. Direct acceleration of a pre-accelerated electron

### 4.1 Simulation Results and Analysis

For convenience we introduce an artificial parameter $D$ that we call the "protracted collision position" and define as the position where the electron would coincide with the pulse peak if the electron were to always travel at its initial speed $v(0)$:

$$\frac{v(0)}{D - z(0)} \equiv \frac{c}{D - z_i} \Rightarrow D \equiv \frac{z(0) - \beta(0) z_i}{1 - \beta(0)} \tag{12}$$

For the initially stationary electron studied in the previous section, $\beta(0) = 0$ so $D = z(0)$ as expected. For values of $D$ far enough from the laser focus such that the electron always experiences a negligibly small electric field (resulting in little change in the electron's velocity from its initial value), $D$ approximates the actual position where electron and pulse peak coincide, hence our name for it. In general, however, the position where electron and pulse peak coincide may be very different from $D$. Although $D$ may not have much physical significance, it is useful as it allows us to control two variables, $z(0)$ and $z_i$, simultaneously: After specifying $D$ for a particular simulation, we use Eq. (12) and our knowledge of the electric field profile to determine the set of values $z(0)$ and $z_i$ closest to the focus but such that the effect of the electric field on the electron is initially below an arbitrarily small amount (i.e. the pulse effectively begins infinitely behind the electron, so the electron effectively begins in field-free vacuum). Simply setting $z_i$ to be an arbitrarily large negative number will of course also produce an accurate simulation, but the simulation time will be unnecessarily long.

In Figs. 6 and 7, we plot the maximum energy gain (normalized by $\Delta E_{\lim}$) and the corresponding optimal $D$ (normalized by $z_0$) vs. $P$ with $w_0$ and the electron's initial kinetic energy $E_K(0)$ as parameters. Fig. 6 and Fig. 7 correspond to the case of $\tau = 7.5$ fs and $\tau = 15$ fs respectively. In Fig. 8, we plot the normalized maximum energy gain vs. $E_K(0)$ with $P$ and $w_0$ as parameters for $\tau = 10$ fs. The plots in Figs. 6-8 are obtained by optimizing over $\psi_0$-$D$ space. From these figures, we observe the following trends:

a) Given $\tau$ and $w_0$, $P_{th}$ decreases with increasing $E_K(0)$. Given $\tau$ and $E_K(0)$, $P_{th}$ increases with increasing $w_0$. $P_{th}$ is approximately independent of $\tau$, as in the $v(0) = 0$ case.

b) Given $\tau$, $w_0$ and $P$, there exists an initial kinetic energy threshold $E_{Kth}$ such that negligible energy gain is obtained for $E_K(0) < E_{Kth}$. Given $\tau$ and $w_0$, $E_{Kth}$ decreases with increasing $P$. Given $\tau$ and $P$, $E_{Kth}$ increases with increasing $w_0$. $E_{Kth}$ is approximately independent of $\tau$. Although some of these trends are evident from Fig. 8, they may all be directly inferred from (a), which tells us that $P_{th}$ is a strictly decreasing function of $E_K(0)$ (given $\tau$ and $w_0$ within the parameter space studied).

Note also that if $P_{th}$ is a strictly decreasing function of $E_K(0)$, $P = P_{th}$ if and only if $E_K(0) = E_{Kth}$.

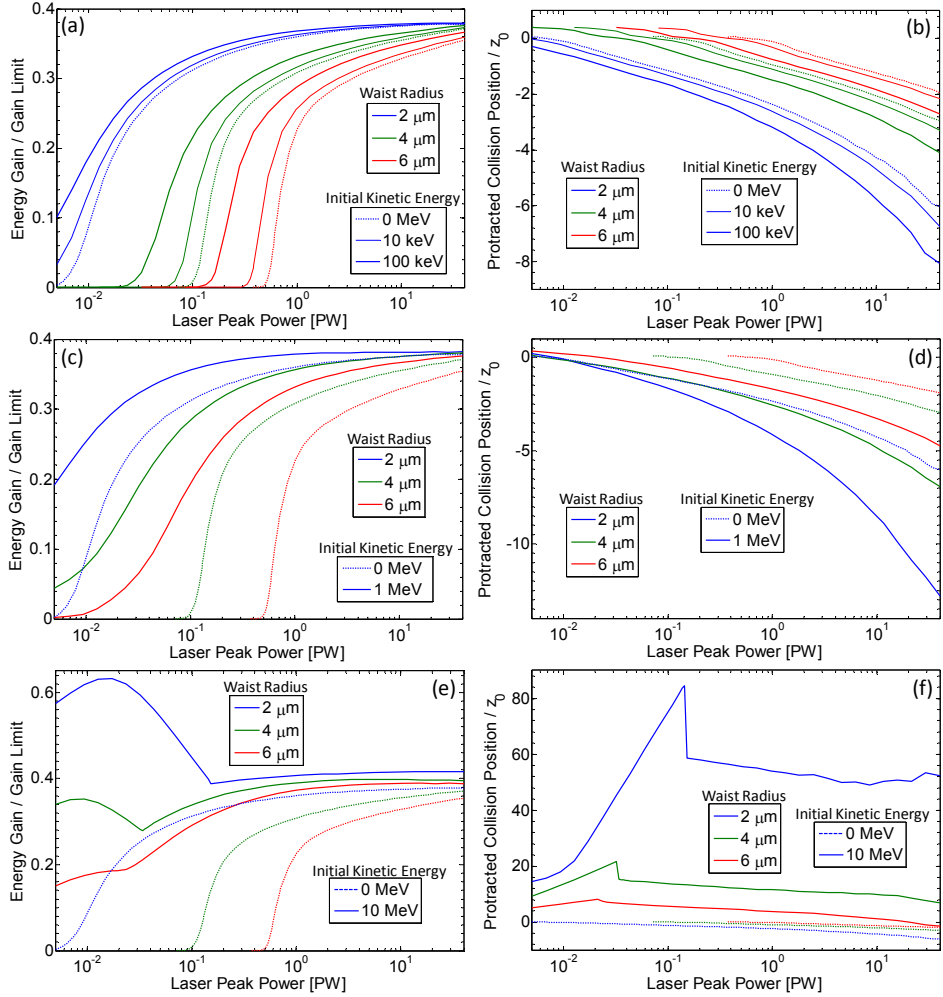

Fig. 6. (Color online) Normalized maximum energy gain and corresponding normalized optimal $D$ vs. $P$ from 5 TW to 40 PW for various $w_0$ and $E_K(0)$: (a), (b) non-relativistic $E_K(0)$; (c), (d) marginally-relativistic $E_K(0)$; and (e), (f) relativistic $E_K(0)$. $\tau = 7.5$ fs. All cases shown correspond to forward scattering of the electron. Cases of very non-relativistic final kinetic energy are not plotted to reduce clutter.

c) Given $\tau$, $w_0$ and $P$, energy gain increases with increasing $E_K(0)$ at least up to a certain $E_K(0)$. As can be seen from the $w_0 = 2$ μm plot in Fig. 8(a), the energy gain starts to fall after a certain $E_K(0)$ (more discussion in Section 4.2).

d) Given $E_K(0)$, $w_0$ and $P$, energy gain increases with increasing $\tau$ up to an optimal $\tau$ and decreases as $\tau$ increases further. As the given $P$ ($E_K(0)$) decreases toward $P_{th}$ ($E_{Kth}$), this optimal $\tau$ increases, showing that longer pulses are favored at lower powers.

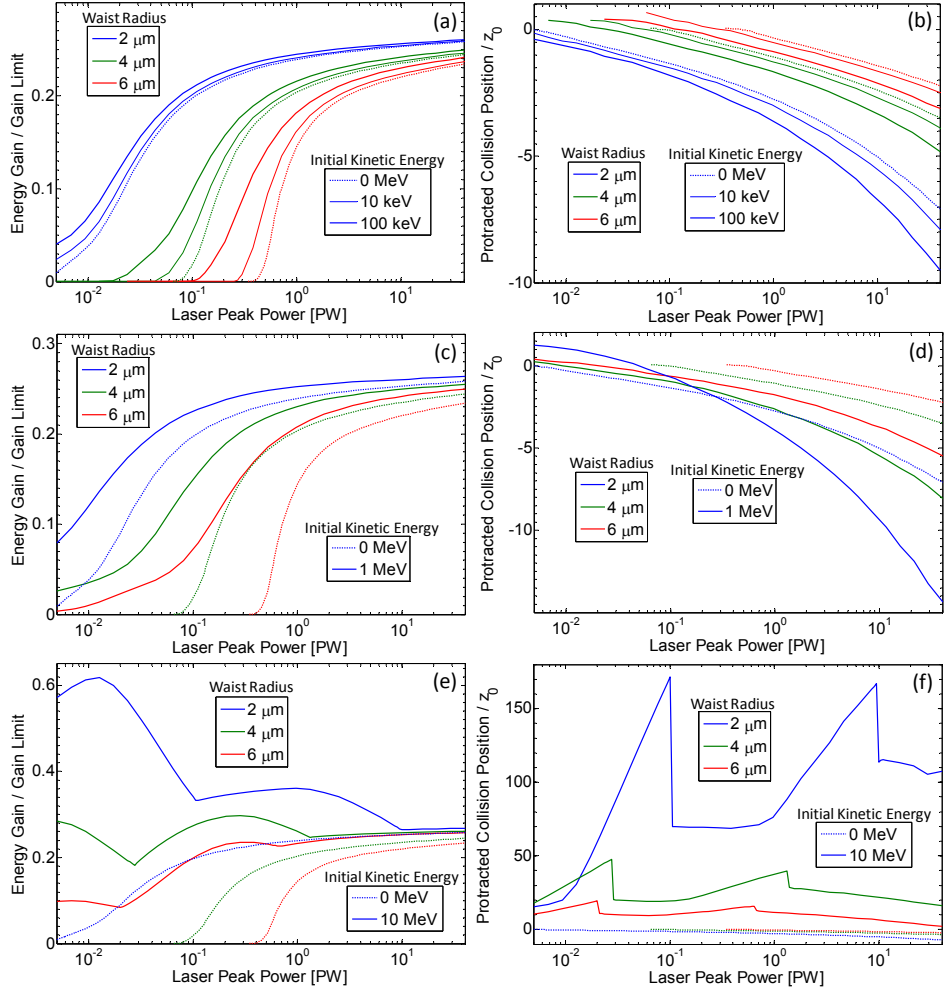

Fig. 7. (Color online) Normalized maximum energy gain and corresponding normalized optimal $D$ vs. $P$ from 5 TW to 40 PW for various $w_0$ and $E_K(0)$: (a), (b) non-relativistic $E_K(0)$; (c), (d) marginally-relativistic $E_K(0)$; and (e), (f) very relativistic $E_K(0)$. $\tau = 15$ fs. All cases shown correspond to forward scattering of the electron. Cases of very non-relativistic final kinetic energy are not plotted to reduce clutter.

e) Given $E_K(0)$, $\tau$ and $P$, energy gain decreases with increasing $w_0$. Once again, the optimal focusing for direct electron acceleration is the tightest as far as we can determine in the paraxial wave approximation.

f) Given $E_K(0)$, $\tau$ and $w_0$, the energy gain in MeV increases with increasing $P$. The energy gain normalized by $\Delta E_{lim}$ also increases with increasing $P$ at non-relativistic $E_K(0)$, but this is not true in general at relativistic $E_K(0)$, as is evident from Figs. 6(e) and 7(e). Fig 9 corroborates our conclusion by showing that the normalized energy gain increases with increasing $P$ for values of $E_K(0)$ up to a few MeV, but ceases to always do so beyond this range. Hence, although greater energy gain in MeV can always be achieved (for given $E_K(0)$, $\tau$ and $w_0$) by increasing $P$ and optimizing parameters, the fraction of the theoretical energy gain limit extracted may in fact become smaller if $E_K(0)$ is relativistic.

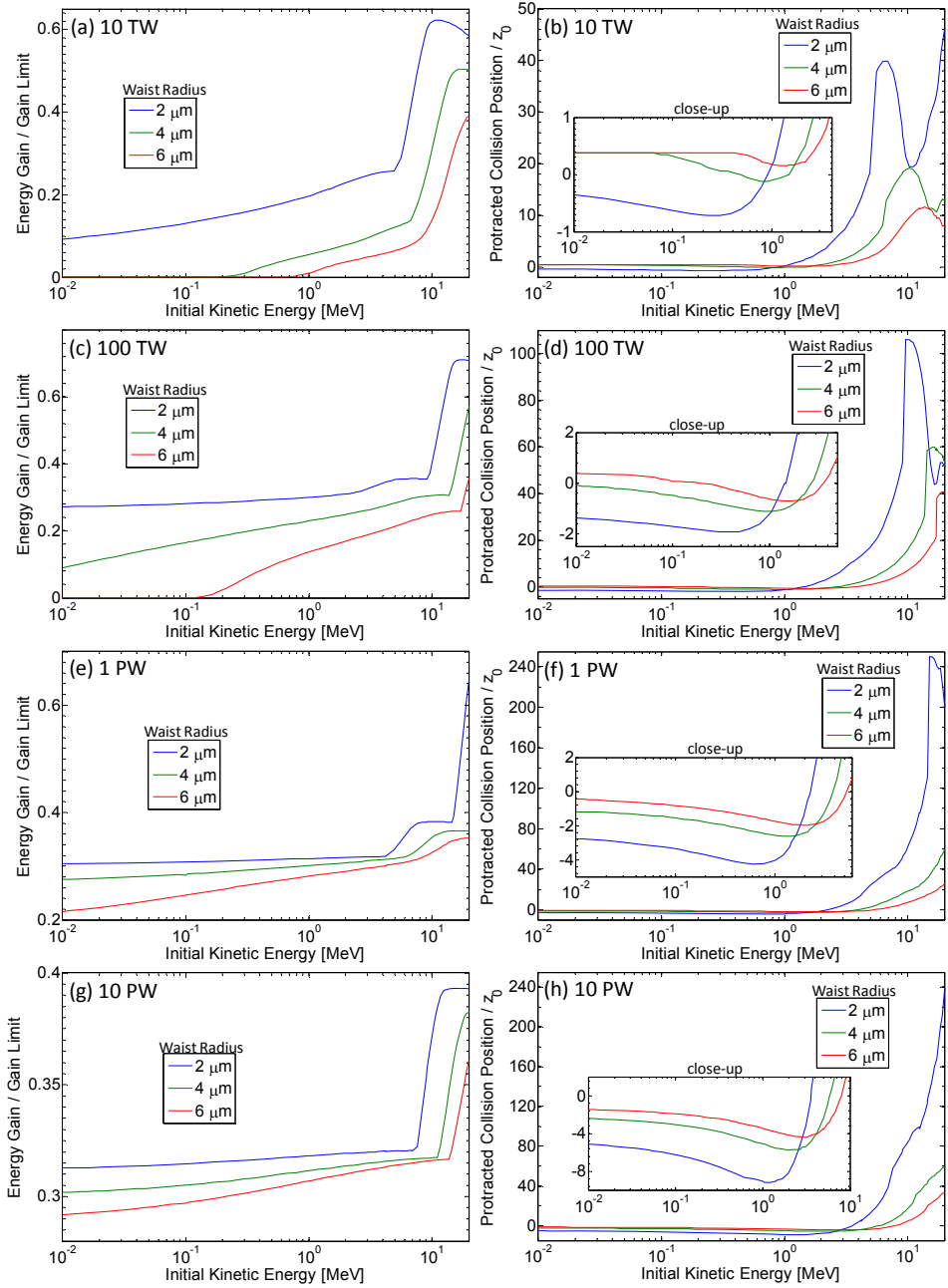

Fig. 8. (Color online) Normalized maximum energy gain and corresponding normalized optimal $D$ vs. $E_K(0)$ from 0.01 to 20 MeV for various $w_0$ and $P$: (a), (b) $P$ =10 TW; (c), (d) $P$ = 100 TW; (e), (f) $P$ = 1 PW; and (g), (h) $P$ = 10 PW. $\tau$ = 10 fs. All cases shown correspond to forward scattering of the electron.

g) At non-relativistic $E_K(0)$, $D$ decreases from its value for the $v(0) = 0$ case with increasing $E_K(0)$. That this decrease is small accords with physical intuition because relative to the speed of the pulse ($c$), an electron with non-relativistic $E_K(0)$ is practically stationary so one would expect the optimal $D$ to be very close to that for the $v(0) = 0$ case. This reasoning, of course, no longer applies at relativistic $E_K(0)$. It is evident from the plots of $D$ in Fig. 8 that beyond a certain $E_K(0)$ (around 1 MeV) for each plot, the slope of $D$ with respect to $E_K(0)$ is no longer always negative, and $D$ itself may be located up to hundreds of times the Rayleigh range beyond the laser focus.

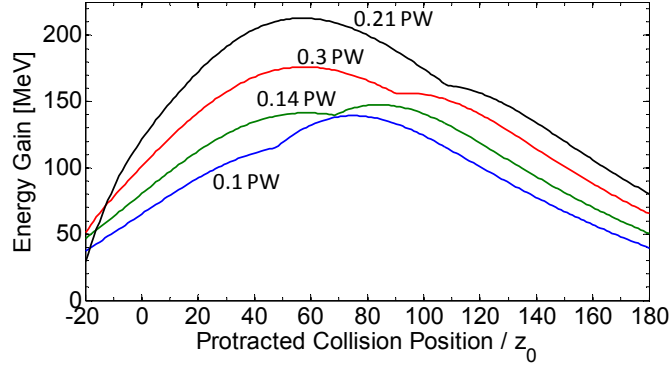

Fig. 9. (Color online) Maximum energy gain vs. normalized $D$ for $\tau = 7.5$ fs, $w_0 = 2$ μm and $E_K(0) = 10$ MeV for various $P$. All cases shown correspond to forward scattering of the electron.

The discontinuities in Figs. 6(f) and 7(f) are due to the existence of multiple energy gain local maxima in $D$ for certain combinations of $E_K(0)$, $\tau$, $P$ and $w_0$. The cause of the discontinuity around $P = 0.15$ PW for the $w_0 = 2$ μm case of Fig. 6(f) is illustrated in Fig. 9, which plots energy gain, maximized over $\psi_0$ space, as a function of $D$. Although each local maxima varies continuously as P increases from 0.14 PW to 0.3 PW, the global maximum jumps at some point from one of the local maxima to the other, resulting in the discontinuity in Fig. 6(f). Similar situations are responsible for the discontinuities in Fig. 7(f).

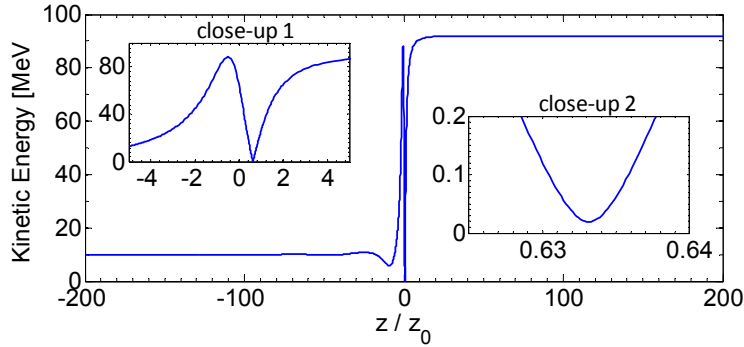

Fig. 10. (Color online) Electron kinetic energy vs. normalized position for $P = 17.3$ TW, $w_0 = 2$ μm, $\tau = 7.5$ fs, $E_K(0) = 10$ MeV, and optimal $\psi_0$ and $D$. Inset "close-up 2" zooms into the point at which the electron enters its effectively final accelerating cycle.

As we have noted, a pre-accelerated electron can gain more than half the theoretical energy gain limit. It does so by entering its effectively final accelerating cycle within a Rayleigh range after passing the laser focus. Fig. 10 shows a plot of kinetic energy vs. displacement for one of the cases picked from the $w_0 = 2$ μm, $E_K(0) = 10$ MeV curve in Fig. 6. As we can see, the electron coming in from the left enters its effectively final accelerating cycle with a kinetic energy of a few tens of keV at a displacement of about $z = 0.633 z_0 < z_0$, and leaves the interaction region with a final kinetic energy of over 90 MeV. The energy gain of over 80 MeV is clearly more than half the theoretical gain limit, which in this case ($P = 17.3$ TW) is about 129 MeV by Eq. (10).

To give an example of how relatively low-power lasers may be used in a direct acceleration scheme, we see that for either $\tau = 7.5$ fs (Fig. 6(e)) or $\tau = 15$ fs (Fig. 7(e)), a pulsed radially-polarized laser beam of $w_0 = 2$ μm and $P = 5$ TW can accelerate an electron from an initial kinetic energy of 10 MeV to a final kinetic energy of about 50 MeV. Eqs. (4) and (5) give us pulse energies of about 45 mJ and 90 mJ for the 7.5 fs and 15 fs pulse respectively. This shows that lasers can already be very useful for electron acceleration at relatively low powers, just that the electrons must be pre-accelerated (preferably to relativistic speeds) to extract substantial energy from the laser pulse. Although it appears from our results that a smaller improvement in normalized energy gain is obtained with a pre-accelerated electron at higher laser powers, this does not discount the possibility of substantial improvements at these higher powers if we increase $E_K(0)$ to values beyond the range studied.

As another example, we note from Figs. 5 and 8 that a two-stage laser accelerator employing a pulsed radially-polarized laser beam of $w_0 = 2$ μm, $\tau = 10$ fs and $P = 10$ TW (giving a pulse energy of about 120 mJ) in each stage can accelerate an initially stationary electron to a kinetic energy of about 6.3 MeV in the first stage, and thence to a kinetic energy of about 36 MeV in the second stage. Note that the same pulse may be used in both stages, since the pulse transfers a negligible fraction of its energy to the electron in the first stage. Clearly, direct acceleration of electrons to substantial energies in infinite vacuum can in principle be realized without the use of petawatt peak-power laser technology. Lasers with peak powers of a few terawatts are already capable of accelerating electrons to energies of tens of MeV, high enough for applications like the production of hard X-rays via inverse Compton scattering [29]. In addition, recall that we have limited our studies to $w_0 \geq 2$ μm. If energy gain continues to increase with tighter focusing for waist radii below 2 μm, it is likely that much more impressive results (at least in terms of energy gain) may be obtained with lasers focused down to an order of a wavelength.

*4.2 Validity of the $v(t) \approx c \: \forall t$ approximation*

Because the electron in Fig. 10 moves at a relativistic speed for most of its trajectory, one may mistakenly expect its energy gain to be approximately 0. This is supported by the egregious approximation that $v(t) \approx c \: \forall t$, which enables an analytic computation of energy gain as (allowing $\alpha$ to be some constant determined by the particle's location relative to the center of the pulse envelope)

$$\Delta E_{v=c} = \int_{-\infty}^{\infty} dz (eE_z) \approx \int_{-\infty}^{\infty} dz \alpha \left[ \frac{e/z_0}{1 + (z/z_0)^2} \sqrt{\frac{8\eta_0 P}{\pi}} \right] \sin\left(2 \tan^{-1}(z/z_0) + \psi_0\right) = 0 \qquad (13)$$

Our exact numerical simulations reveal that this is not the case. Although the electron is relativistic for most of its trajectory, the few places at which it becomes non-relativistic are sufficient to produce an asymmetry that prevents the actual integral of force over distance from vanishing.

This observation also encourages the hypothesis that the highest $E_K(0)$ with which an electron may be substantially accelerated by a pulsed radially-polarized laser beam is on the order of the theoretical gain limit $\Delta E_{\lim}$, because $\Delta E_{\lim}$ also represents the maximum *deceleration* of a pre-accelerated electron. If $E_K(0)$ is relativistic and $E_K(0) >> \Delta E_{\lim}$, the laser field can never at any point decelerate the electron to non-relativistic speeds so $v(t) \approx c\ \forall t$ would be true and Eq. (13) would hold. This hypothesis may be extended to any other direct acceleration scheme if a corresponding $\Delta E_{\lim}$ expression may be found for it. The electron's energy gain for a given laser should thus decrease after some point as $E_K(0)$ continues to increase, and become negligible for $E_K(0) >> \Delta E_{\lim}$. This implies that there exists a second set of power and initial kinetic energy threshold values (i.e.: different from $P_{th}$ and $E_{Kth}$ in Section 4.1) observable only at values of $E_K(0)$ beyond the range studied given our range of $P$. This second threshold places an upper bound on $E_K(0)$ given $P$ (and continues to place a lower bound on $P$ given $E_K(0)$) for non-negligible acceleration.

## 5. Conclusion

We have studied the direct acceleration of a free electron in infinite vacuum along the axis of a pulsed radially-polarized laser beam. By introducing appropriate normalizations to the electrodynamic equations, we have shown that our results for $\lambda = 0.8\,\mu m$ may be readily scaled to obtain the results for any $\lambda$. An important consequence of this is that for a given peak power, a larger pulse energy is required for exactly the same maximum acceleration at a larger $\lambda$ if focusing ($w_0/\lambda$) remains constant, because the number of carrier cycles in the pulse envelope must also remain constant, leading to a longer pulse.

In all cases studied (regardless of power, pulse duration and electron initial speed), the greatest acceleration is achieved with the most tightly-focused laser. Also, the optimal pulse duration is a function of power, with shorter pulses favored at higher powers and longer pulses favored closer to the threshold. In all cases studied, energy gain in MeV increases with increasing peak power, but the energy gain normalized by the theoretical energy gain limit does not always do so. Greater acceleration may be achieved with pre-accelerated electrons. The net energy gained by an initially relativistic electron may even exceed more than half the theoretical energy gain limit, which is not possible with an initially stationary electron in the parameter space studied. We have also given some examples of how electron acceleration by tens of MeV is in principle demonstrable with laser powers as low as a few terawatts. Finally, we have hypothesized that the approximation $v(t) \approx c\ \forall t$ will be valid (and electron acceleration thus negligible) if the electron's initial kinetic energy greatly exceeds the laser's theoretical gain limit. Future studies will examine the impact of various direct acceleration parameters on beam emittance and energy spread of an electron beam.

**Acknowledgments**

This work was financially supported by the National Science Foundation (NSF) grant NSF-018899-001 and the Agency for Science, Technology and Research (A*STAR), Singapore.